\documentclass[twocolumn]{article}

\usepackage{fullpage}
\usepackage{times}
\usepackage{cite} 
\usepackage[breaklinks,colorlinks,linkcolor=blue,citecolor=blue,urlcolor=blue]{hyperref}
\usepackage[hyphenbreaks]{breakurl}
\usepackage{xspace}
\usepackage{color}
\usepackage{graphicx}
\usepackage{epigraph}
\usepackage{subfig}
\usepackage{breakcites}

\long\def\com#1{}

\long\def\xxx#1{}

\newcommand{\ie}{{\em i.e.}\xspace}
\newcommand{\eg}{{\em e.g.}\xspace}

\newcommand{\coin}{{Popcoin}\xspace}
\newcommand{\coins}{{Popcoins}\xspace}
\newcommand{\atom}{{Poplet}\xspace}
\newcommand{\atoms}{{Poplets}\xspace}

\title{Democratic Value and Money for Decentralized Digital Society \\ ~\\
	{\large\em preliminary work-in-progress; may become
		\href{https://bford.info/book/}{part of a future book} \\
	written and first distributed June 2018 
	in discussions for the edited volume \\
	\href{https://press.uchicago.edu/ucp/books/book/chicago/D/bo68657177.html}{Digital Technology and Democratic Theory}
	\\ ~}}
\author{Bryan Ford}
\date{}

\begin{document}
\maketitle

\begin{abstract}
Classical monetary systems
regularly subject the most vulnerable majority of the world's population
to debilitating financial shocks,
and have manifestly allowed uncontrolled global inequality over the long term.
Given these basic failures, how can we avoid asking
whether mainstream macroeconomic principles
are actually compatible with democratic principles such as equality
or the protection of human rights and dignity?
This idea paper takes a constructive look at this question,
by exploring how alternate monetary principles
might result in a form of money more compatible with democratic principles --
dare we call it {\em democratic money}?

In this alternative macroeconomic philosophy,
both the supply of and the demand for money must be rooted in {\em people},
so as to give all people both equal opportunities for economic participation.
Money must be designed around equality,
not only across all people alive at a given moment,
but also across past and future generations of people,
guaranteeing that our descendants cannot be enslaved by their ancestors'
economic luck or misfortune.
Democratic money must reliably give all people a means to enable
everyday commerce, investment, and value creation in good times and bad,
and must impose hard limits on financial inequality.
Democratic money must itself be governed democratically,
and must economically facilitate the needs of citizens in a democracy
for trustworthy and unbiased information
with which to make wise collective decisions.
An intriguing approach to implementing and deploying democratic money
is via a cryptocurrency built on a {\em proof-of-personhood} foundation,
giving each opt-in human participant one equal unit of stake.
Such a cryptocurrency would have both interesting similarities to,
and important differences from, a Universal Basic Income (UBI)
denominated in an existing currency.
\end{abstract}

\tableofcontents

\section{Introduction}

Today's nominally democratic capitalist socioeconomic structures
are failing in numerous critical ways:
failing to engage and empower citizens in democratic processes, 
failing to offer citizens
with trustworthy information sources, 
resistant to data-driven foreign and domestic propaganda campaigns, 
failing to drive political dialog
toward consensus and away from polarized tribalism, 
failing to manage the corrupting influences of wealth
on political decision, 
failing to offer citizens opportunities and economic empowerment
in the face of ubiquitous automation, 
failing to control growing inequality sufficiently
to maintain confidence in the fairness or stability
of the system, 
and
failing to generate the collective wisdom or power of action
to address existential global challenges such as climate change, 
These ongoing failures cast increasingly-widespread doubts
about whether democratic self-rule is up to the task, 
and leading large populations toward following ideologues
with offering quick and easy answers and scapegoats. 

This paper makes the case that the problem is not that
democratic self-rule is inherently flawed or inadequate,
but that the current embodiments of this concept are merely engineered
insecurely and inadequately,
and in fact are {\em not democratic enough}
to offer sufficient security, stability, or equitability
in the conditions of today's digital age.
In particular, a key problem is that the principles of democracy
are traditionally applied only to one tiny slice of society's operation --
namely elections for public office --
leaving other key institutions that democracy ultimately depends on,
such as economic activity and information dissemination,
operating in fundamentally flawed and non-democratic structures.
Democracy is a technology for self-organization,
and like any fundamental technological idea,
there are any infinity of ways
it can be inadequately, insecurely, or incompletely implemented.
If a toaster electrocutes its user or a bridge collapses,
we do not question the fundamental utility of toasters or bridges,
but instead ask what went wrong in their engineering or manufacture
and attempt to make the next toaster or bridge stronger and safer.

This paper attempts to sketch a ground-up, ``first-principles'' reconsideration
of how democratic self-organizing systems can or should be designed,
in light of both the technological capabilities now at our disposal
and the painful lessons we have learned from the many failures of
our prior toasters, bridges, and democracies.
Beyond serving as merely a philosophical thought-experiment,
this redesign sets the goal of leveraging the present-day opportunity
for in-practice ``permissionless innovation'' and deployment,
afforded by the popularization of decentralized systems
such as cryptocurrency, distributed ledger, and blockchain technologies.
In short, while we make no pretense that established democratic systems
would readily or willingly make structural changes
as deep or fundamental as the ones suggested here,
but instead observe that today's decentralized systems technology
is becoming mature enough to give us the tools to ``grow''
a next-generation system of democratic system of self-organization
gracefully alongside the existing gradually-failing ones,
coexisting peacefully in the technological innovation space
without requiring either the support or even explicit permission
of the currently-prevailing authorities.

The focus of this paper is primarily on identifying
a set of guiding design principles for next-generation,
technology-driven systems for democratic self-organization.
An important secondary goal is to identify particular
implementation challenges and security risks,
and to sketch and point to potential solutions to them.
With this in mind,
we first outline the general principles and goals
we wish to be embodied pervasively in a self-organizing system design,
then subsequent sections will focus on the key operational components
of information-gathering, decision-making, and incentives/economics.

Everything in this paper should be considered a work-in-progress:
the proposed principles are probably neither complete
nor as well-formulated as they could be,
and the technical approaches suggested are merely examples
with no claim that they represent the best possible design.
But then, any healthy democracy must also consider itself
always a work-in-progress as well,
evolving to meet new challenges and fix flaws and vulnerabilities,
as we will explore later in Section~\ref{sec:evo}.

\subsection{Summary of Core Principles}
\label{sec:intro-principles}

We propose that a truly democratic self-organizing and self-governing community
must embody the following principles pervasively
throughout its design and operation:
\begin{enumerate}
\item	{\bf People as the foundation of power and value:}
	Democratic power and wealth ultimately originates in people (humans),
	regardless of how it might subsequently flow.
	Similarly, any legitimate democratic measure of value --
	of ideas, candidates, property, services, {\em anything} --
	must be founded in the value of these things {\em to people}.
\item	{\bf Equal opportunity over population:}
	All humans living at any given time must have
	an equal right to participate
	and must wield an equal fundamental power share
	in all self-organization processes,
	regardless of individual characteristics
	such as age, gender, skin color, etc.
	Only people (humans) enjoy this fundamental right of equal opportunity:
	legally or technologically constructed entities
	such as corporations and botnets do not.
\item	{\bf Equal opportunity over time:}
	All humans living at any time must have
	equal participation rights and fundamental power share
	with respect to their predecessors and ancestors.
	The successes and mistakes of prior generations cannot
	dominate the power, opportunity, or maneuvering room
	of their descendants.
\item	{\bf Democratic governance:}
	All decisions governing the community's operation and evolution
	are made through democratic deliberation and decision processes
	ensuring that no one is disenfranchised from the right to participate --
	including those who for innumerable reasons
	may have limited time or ability to participate.
\item	{\bf Democratic economy:}
	Participants must have a viable basis for everyday commerce,
	investment, value measurement, and value creation
	that is compatible with the above democratic principles of equality.
	This value basis must not devolve into economic aristocracy over time
	through the gradual confinement of individual and collective power
	to ever-smaller shares
	of the world's total wealth and economic power.
\item	{\bf Democratic information:}
	The processes of gathering, filtering, curating, and disseminating
	the information needed for well-informed democratic self-governance
	must be likewise secured through democratic processes,
	ensuring that no internal or external actors can obtain and wield
	democratically disproportionate influence
	on the perceptions, decisions and collective actions of the community.
\end{enumerate}

This paper focuses for now primarily on the problem of
rethinking and redesigning the economic notions of value and money
for consistency with democratic principles,
while only briefly touching on the equally-important problems
of strengthening the democratic foundations and processes
for governance and information-gathering.
This initial proposal is thus intended to represent only a seed
to be expanded in the coming months (and perhaps years)
into a more complete theory of democratic self-organization in the digital age.

We therefore first develop a democratic notion of value
in Section~\ref{sec:value},
then build on this a concept of democratic money
in Section~\ref{sec:money},
and explore its application to wealth and property ownership
in Section~\ref{sec:property}.

{\em Note: only Sections 1--\ref{sec:money} written so far;
the others may be relegated to separate papers,
and the above intro will need to be narrowed accordingly.}

\xxx{ Idea to work in: Democratic Decantralized Autonomous Organizations, DDAO:
	enforces property that all human stakeholders have an equal
	share of governance power over the organization,
	scaling in power from 0\% to 100\% as the company's reach
	grows from a tiny company to a global company that impacts everyone. }

\section{Democratic Measures of Value}
\label{sec:value}

\epigraph{Man is the measure of all things.}{Protagoras}

Money, in the form of innumerable types of currencies,
are a technology that for millenia have served as a society's
primary mechanism to measure value,
as well as to incentivize and reward production of value,
including both physical goods and non-physical services and information.
But while money and its capitalistic uses
have rightfully held a culturally-accepted central role
in the evolution of democracy and political theory,
the principles underlying money and finance have largely followed
a separate, parallel track from politics and democracy:
separate technologies playing different roles by vastly different rules,
though with well-studied, complex, often problematic interactions.
But is this strong separation between the theories of democracy and finance
fundamentally either necessary or desirable?

\subsection{People-Centric versus Thing-Centric Value Bases}

We explore here whether more tightly intertwining fundamental principles
of democratic and economic theory in the right fashion
could substantially benefit both,
each helping to address traditional limitations in the other.
Democracy is founded on a people-centric foundation of value:
that all citizens or eligible voters
should have equal rights and equal voting power,
at least as a starting point in a process of political decision-making,
and that those who obtain and exercise unequal power in this process
(\eg, elected officials and the bureaucracies they supervise)
should ultimately serve and be accountable
to the entire citizenry as a collective.
The accepted democratic measure of the ``value''
of a candidate or party platform
is based on the number of citizens or voters supporting them.

Money, by contrast, traditionally embodies a
{\em thing-centric} foundation of value,
originally based on the scarcity of materials such as beads or precious metals,
and gradually transitioning to a state-managed notion of scarcity and value --
a transition that Nixon completed with the abolishment of the gold standard.
Even after this transition to fiat currency,
whose value ultimately depends entirely on collective belief in its value,
money is still viewed, measured, and analyzed in terms of the {\em things}
that money can buy or assist in manufacturing through capitalistic investment.

While the people-centrism of democracy's value basis is still
widely and rightfully accepted as a social good,
the flow of political power in today's democracies
occurs in extremely coarse-grained, illiquid units
enabled by periodic elections,
which are in practice so disconnected from citizens' daily lives
that elections often fail to persuade even half of the citizenry
that participating is worth their time.
On the other hand,
money is extremely liquid and in ubiquitous use all the time,
especially in its modern electronically-facilitated forms,
but the utility and legitimacy of its thing-centric basis of value
is increasingly questionable in an age where automation
rapidly decreases both the scarcity and hence value basis of physical goods
and human labor in creating them,
and digitalization and AI similarly undermines
the scarcity and hence the value basis of white-collar
human labor in administrative and professional services.

In short, democracy gets the people-centric value basis right,
but its utility and relevance is undermined
by its clunky and illiquid dependence on periodic elections,
while money is liquid and ubiquitous but its thing-centric value basis
but appears to be increasingly disconnected
from the needs of the people using it.
Can we combine the most important elements of each
to create useful ``currencies'' -- measures of value --
that embody both the people-centric value basis central to democracy
together with the liquidity and utility needed to make them
useful to people all the time in their everyday lives?
We next sketch one potential approach to answering this question.

\subsection{Traditional Elections as an Ephemeral Democratic Currency}
\label{sec:election-currency}


In economics,
{\em money} can in principle be anything that generally serves
as a medium of exchange, a unit of account, and a store of value. \xxx{citation}
Government-issued currencies,
designed to be ``legal tender'' within a given jurisdiction,
are obviously the most commonly-used form of money today.
But other things -- such as gold or Bitcoin --
can arguably serve as money to varying degrees
depending on the extent to which they satisfy these properties of money.

Taking the definitions of money and currency broadly,
we may view traditional democratic elections
as a highly constrained and ephemeral form of currency.
An election in effect issues each voter a single quantum of a currency
that springs into existence for that election alone,
and ceases to exist in any useful form after the election is concluded.
The government creates this democratic currency
by issuing each voter a single currency unit,
taking the form of a ballot instead of a banknote,
which the voter may then ``invest'' by casting it
for a candidate or position of his or her choice.
This ephemeral currency acts as a unit of account
because cast votes serve as an explicit, quantified measure
of the {\em democratic value} a given candidate or choice has 
to the voting population.

The utility of this ephemeral currency as a medium of exchange
is extremely limited, of course:
it is generally unaccepted and often illegal to trade votes 
for money, goods, or services.
Instead, votes are traditionally ``tradeable'' for only one thing:
the {\em chance}, however uncertain it may be to materialize,
of placing a preferred candidate into power
or deciding an issue in the way the voter prefers.
Finally, votes serve as an extremely short-lived store of value,
namely from the time they are cast
until the time the election's results are decided and formally announced.

Despite these considerable restrictions,
we can argue that elections already inherently and essentially share
key properties of currency,
being most importantly a democratic measure of value or ``unit of account,''
secondarily a medium of exchange (of votes for a chance at influencing power),
and finally an extremely short-lived store of value during the election.
Accepting this connection between elections and currencies
enables us to explore more deeply whether and how it might be useful
to bring democratic measures of value --
in the ``one person one vote'' sense --
into the traditional realm of currencies.
We also wish to explore on the other hand how to bring more of
the useful properties of traditional currencies --
namely less-restricted utility as a medium of exchange
and less ephemeral store of value --
into the realm of democratic self-organization and decision-making.
Can we design democratic currencies
retaining the fundamental egalitarian value basis of elections
while providing greater liquidity and utility to make flows
of {\em democratic value} more explicit, transparent, fair, and sustainable?

\subsection{More Liquid Foundations for Democratic Value}

A conventional election typically serves the purpose
of deciding {\em one particular} question at a time,
independent of all other decisions --
such as which candidate(s) to elect to a certain governmental role
or whether to approve a particular initiative.
This decision-making structure builds on the implicit premise
that by some {\em other} means it has been decided that
it is time to hold elections to fill these particular positions
or to involve the voters in a particular decision:
\eg, because the constition demands such elections at a given frequency,
because a current government has ceased to function and needs a fresh mandate,
or because a threshold of voters have signed a petition
to bring an initiative or referendum up for popular vote.
These are all different, and all relatively illiquid,
approaches to {\em prioritization}:
determining what activities or issues are worth spending
the voter's and government's time and resources on.

There are precedents for more liquid, currency-like approaches
to democratic prioritization, however.
One rather common and intuitive such approach is {\em cumulative voting},
in which each voter is given more than one vote -- but still an equal number --
to allocate to available candidates, positions, or priorities as they see fit.
Suppose an election is held whose purpose is to prioritize the allocation
of the community's time, funding, and/or other resources
among several perceived problems to be addressed or directions for development:
\eg, between investing in education ($A$), environment ($B$), or economy ($C$).
There may be no question that all these are important problems
worthy of attention;
the question is only one of prioritizing
and allocating scarce resources between them.
One classic approach to democratic prioritization
is to give each voter, say, 10 votes to allocate as they prefer.
Thus, a voter who cares most about eduction, a little about environment,
and not at all about the economy might cast six votes for choice $A$,
four for $B$, and none for $C$.
There is of course plenty of room for debate --
both in electoral theory and in practice --
on whether and in what contexts cumulative voting is a ``good'' way
to make prioritization decisions such as this.

For our purposes the point is that cumulative voting offers
one intuitive, widely-used, and easily-understdood precedent
for more liquid, currency-like approaches
to democratic prioritization and decision-making.
And if it is democratically legitimate
to give each person 10 votes in a prioritization decision,
then is there any fundamental reason we could or should not
make such prioritization decisions even more fine-grained and liquid,
for example by giving each person 100, 1000, or 10,000 votes,
as long as votes are equally distributed?
Subdividing voting power into more fine-grained units
in principle enables people to express more subtle preferences
across a wider range of potential choices,
and makes it possible to assign quantifiable measures of democratic value
not only among some preselected ``top few'' choices
but instead perhaps among a relatively open-ended set of choices
that anyone might propose.

\xxx{ note: cumulative voting is definitely not the only
	and probably not the best way to handle participatory
	budget decisions in general;
	for more recent alternatives see for example
	Knapsack Voting~\cite{XXX}.
}

\xxx{ see also quadratic voting in posner15voting:
"In a race with more than two candidates, or an election where more than a
single issue is decided, individuals could be allocated an artificial currency
that they could use to quadratically buy votes on individual issues."
}

\subsection{Time-based Foundations for Democratic Currencies}

Both supply-side and the demand-side factors in the thing-based economy
tend to vary drastically over time.
The currently-fashionable mix of goods people demand,
the prices of those goods,
the costs to produce and distribute them,
the human labor involved in this production
that is inexorably giving way to automation,
all amount to different flavors of quicksand
if we seek any stable (let alone democratic) measure of value.

Many have observed, however, that {\em time itself} --
or specifically the time of any given person --
might offer a fundamentally more stable and democratic basis
for measuring value.
All people inherently have access to the same universal supply of time,
which we all consume at approximately the same rate,
ignoring for now the plight of people traveling close to the speed of light.
For this reason,
person-time has significant appeal as a democratic measure of value:
a measure that can be said to be just as ``fair'' as votes
in the sense of equality of power and opportunity over population,
but significantly more fine-grained and liquid than conventional votes.

This observation has formed the basis of time-based currencies globally,
such as that first suggested and put into practice
in Japan by Teruko Mizushima~\cite{miller08teruko},
in which people can bank and trade hours of effort helping each other.
Analogous time-based community currencies have been explored
in the US and elsewhere,
such as
Time Dollars~\cite{cahn92time},
Ithaca HOURS~\cite{glover95hometown,hermann06special,jacob12social},
and most recently in highly-experimental cryptocurrencies
such as Nimses~\cite{nimses17,larkina17nimses}.

\com{
From hermann06special:
HOURS are further circumscribed by their very purpose:
facilitating local exchange.
Hence, HOURS are also defined by where they do not flow,
most notably into the coffers of national and international chain stores.
}

Mizushima's conception of time-banking
held to a strongly-egalitarian principle
that everyone's time has equal value.
This principle may help foster community spirit
and may be practical where the type of work being traded
and the level of expertise required for that work is relatively homogeneous,
as in the caring-centric work that Mizushima's time-bank focused on.
Other time-banking systems such as Ithaca HOURS, however,
have adapted a more weakly-egalitarian model 
in which one person-hour of currency is intended only
to be a nominal or average guideline for pricing and trading services,
recognizing and explicitly accepting that skilled professionals
might reasonably charge several HOURS per hour of their highly-demanded time.

\com{	Time-based value basis precedent
	Teruko Mizushima
	Ithaca HOURS
	Cahn, "Time Dollars"
	Nim currency: 1 minute of life

	Teruko Mizushima~\cite{miller08teruko} created in 1973
}

\section{Democratic Money and Currency}
\label{sec:money}

Having reviewed several precedents for more democratic,
people-focused notions of value,
we now turn toward developing a notion of money
intended to serve in the traditional roles of money --
namely as a unit of account, a medium of exchange, and a store of value --
but ideally grounded more firmly on democratic values and principles.
We then explore technological means by which we might establish
{\em democratic currencies} that could make democratic money
explicit, tangible, and useful in practice.

\subsection{Key Principles for Democratic Money}
\label{sec:money-principles}

We first briefly summarize several key principles
we would like a democratic currency to embody,
then subsequently delve further into these goals
and potential ways to achieve them.
These principles intentially reflect
the broad principles proposed earlier in Section~\ref{sec:intro-principles},
but focused specifically on their application to democratic money.
\begin{itemize}
\item	{\bf People-centric value basis:}
	A democratic currency assigns value to goods and services
	only to the extent that they are demanded {\em by people}.
	Constructs such as corporations can only transmit, not create,
	real value.
\item	{\bf People-centric money supply:}
	Just as democratic political power
	originates in people
	and flows into representative government via elections,
	economic power in a democratic currency
	originates in people and flows into the economy via money.
	Only people create money;
	governments and banks are at most tools in this process.
\item	{\bf Equality over population:}
	Just as each voter in an election wields equal voting power,
	each participant in a democratic currency
	must have an equal share of fundamental economic opportunity
	with respect to peers.
	\com{
	This means both an equal share in
	defining the value of goods and services through demand.
	}
\item	{\bf Equality over time:}
	Participants in a democratic economy at any given time must must have
	the same fundamental economic power and opportunity
	as those coming before and after.
	In particular, one generation's economic opportunity must not be
	dominated by the winnders and losers of prior generations.
\item	{\bf Entrepreneurship:}
	Within the above constraints,
	all participants have equal opportunity to benefit and profit
	through excellence, innovation, wise investment,
	thereby potentially becoming unequal in outcome.
\item	{\bf Stability:}
	The value and purchasing power of the currency
	should be stable and not swing too rapidly, widely, or unpredictably,
	at least with respect to essential commodities such as food and clothes
	that themselves embody moderately-stable needs of people.
\end{itemize}

The bulk of traditional economic research and innovation
for the past century has been dominated by the singular goal of {\em stability},
attempting to fix the perennial boom-and-bust cycles of capitalism.
While stability is undoubtably important,
we propose that stability should be only one of several fundamental goals
that we should be designing our money and economic theories around.
We might wonder, in fact,
whether in its nearly-single-minded focus on stability as the sole end-goal,
traditional macroeconomics could be falling into a trap analogous
to the beginning driver who focuses on the car immediately ahead,
or the beginning pilot chasing a desired altimeter and airspeed reading,
resulting in a wobbly and potentially dangerous ride.
Both the seasoned driver and the experienced pilot know
to keep their attention {\em on the horizon} -- where they're going --
and use their instruments only to calibrate and confirm
the details of their long-distance trajectory.
Could our inability to achieve economic stability be in part
be in part precisely because we are endlessly preoccupied with
reactively chasing short-term economic instrument readings
without having sufficiently meaningful far-horizon goals to keep our eyes on?

\xxx{
A standard rule of driving is to keep one's eyes on the horizon --
where you want to go --
and not on the road or car immediately ahead or on the steering wheel.
Pilots must learn a similar rule of flying ``straight and level'':
to keep one's eyes primarily on the horizon and attitude,
using other instruments only to confirm accurate flight:
``chasing'' a target altitude or airspeed is a recipe for a wobbly flight.
But in pursuing stability as the primary goal,
and reactively chasing corrections to immediate economic indicators
instead of some long-term ``horizon'' goal,
are today's central bankers falling into the same trap
as the beginning driver or pilot?
But if we would like to pursue a far-horizon goal,
what should that goal even be?
}

\subsection{A Reference Design for Democratic Money}

Given the above basic goals,
we now briefly summarize a {\em reference design} for democratic money,
intended to satisfy the combination of the above principles,
and to be potentially implementable
in the form of a modern permissionless cryptocurrency we call \coin.
We then subsequently elaborate the reasoning behind this reference design.
\begin{enumerate}
\item	Whenever new currency is created,
	it is distributed not to banks
	but directly to all human participants as a {\em basic income},
	in equal portions,
	to ensure equality of opportunity over population
	in foundational purchasing power.
\item	Democratic currency represents a {\em limited-term} power
	to spend, invest,
	and enjoy the rewards of fruitful investment in one's lifetime,
	but not an aristocratic right to economic power
	to be passed across generations.
	Currency therefore has a nominal lifespan of 50 years,
	calibrated to approximate the working lifespan of a modern human.
\com{
\item	Because existing \coin devalues at a rate of 5\% per year,
	newly-created \coin has a nominal lifespan of 20 years,
	calibrated to enable its limited accumulation
	for employment in large ventures,
	while ensuring a strong incentive
	for productive investment rather than holding of free currency.
}
\item	At any time there is a {\em value space} of finite size
	representing the sum total of all money that exists or could exist.
	\xxx{ Reflecting the fact that the planet and its resources
		and carrying capacity are finite.
		See Kenneth Boulding 1973 quote. 
		A closed-world monetary model for a closed-world planet.}
\item	Each year, all existing currency is first devalued
	by the reciprocal of its nominal lifespan,
	\ie, by one-fiftieth of its current value in the reference design.
	New money is then created to fill exactly the corresponding portion
	(\ie, 1/50) of the total value space
	and distributed in equal measure among all human participants.
\item	Because these distributions are the {\em only} way \coin is created,
	the basic income's purchasing power is defined
	not by policy but implicitly as a share of the total useful utility
	the currency is providing its users collectively.
\item	This constant-rate devaluation and distribution ensure
	a stable balance between scarcity and a renewing supply of money,
	ensure equal opportunity over population (within a given year),
	ensure equal opportunity over time (across years and generations),
	and incentivize the circulation of money for productive use.
\item	At any time we define the value of one \coin to be 
	the size of the total value space,
	divided by the number of participants in the most recent distribution,
	divided by the average number of days per year (365.25).
	One \coin is thus a stable representation of a one-day share
	of one person's basic income.
\item	When participation rate increases faster than devaluation,
	participants in earlier distributions receive a larger slice
	of that distribution,
	since each distribution divides a fixed portion
	of total value space.
	These larger slices reward early adopters
	and incentivize participants to join early and to sign up others.
	This early adopters reward is transparent and self-limiting, however,
	tapering off smoothly as participation approaches total population.
\end{enumerate}

We now explore and further develop the principles for democratic money,
and how the reference design satisfies them.


\com{
A contrasting, modern Libertarian perspective essentially proposing a return to the gold standard with a 100

In a society with a pure gold standard and a 100-percent
reserve requirement, all citizens would gain from the gradual,
continuous increase in the purchasing power of their monetary
units. p.765

- a great example of distribution-blind
"rising tide raises all boats" logic,
which ignores the fact that even if this is so,
for many non-monetary reasons the haves will keep increasing their share,
economically crowding out the have-nots in the total wealth pie.
}

\subsection{Equal Opportunity over Population in a Democratic Currency}
\label{sec:eq-pop}

Perhaps the most fundamental principle of democracy is that
all first-class participants or citizens are presumed
to wield an equal share of fundamental power in the collective (\eg, one vote),
even if varying choices can lead to different effective power outcomes
(\eg, serving in public office versus remaining a private citizen).

Reflecting this basic principle,
we define a {\em democratic currency}
as a measure of value that has most of the liquidity of traditional money,
such as fungibility and divisibility,
but is founded on a people-centric rather than thing-centric value basis.
In a democratic currency, value, like voting power,
ultimately originates from all the individual participants
in equal portion, ``one person one vote.''
In a democratic currency, whenever and however new money is created,
it is not distributed to a centralized hierarchy of banks,
who further loan it out to people or businesses the banks judge deserving.
Instead, newly-created money is distributed directly
to all the individual participants (citizens or voters), in equal share,
at the time of creation.

After creation and initial distribution in this democratic fashion, however,
individuals are free to trade this currency for goods or services,
invest it in anticipation of a future profit,
or employ it for any of the traditional uses money serves.
This use of the currency can of course produce winners and losers,
and indeed should,
to incentivize participants to use and invest their resources carefully,
and to reward those who produce value
in other forms tradeable with the currency.
Ensuring that participants have both equal opportunity or starting position,
and equal ``opportunity to become unequal'' through subsequent action,
is already thoroughly accepted not only in capitalist economics
but in the practical operation of modern democracies,
which offer privileged and decidedly unequal levels of power
to reward political candidates who successfully convince voters
of the value of their abilities or platforms.

\xxx{ point out, briefly discuss close relationship
	to ideas of delegative/liquid democracy}

\subsection{When and How Often to Create Democratic Money?}

As a naive starting point,
we could consider creating a democratic currency in a ``one-shot'' fashion.
Like the ephemeral currency that a traditional election effectively creates,
as discussed earlier in Section~\ref{sec:election-currency},
the government could at some particular moment
launch a brand-new currency by printing a certain number of new banknotes
and distributing an equal share to each citizen.
Unlike votes in an election, however,
citizens would be free and encouraged
to trade these new banknotes for goods and services,
and/or to use them as a short- or long-term store of value.
These banknotes would take on some value based on their scarcity, utility,
and the population's confidence in the government create them
(\eg, trust in the government not to cheat or devalue the currency
by secretly printing and distributing more banknotes than should exist
according to the total population and fair-share amount per person).
\xxx{	mention that this actually happened in Russia's tradition
	to capitalism; ask Anya for references. }

Such a ``one-shot'' currency distribution could obviously be said
to be ``fair'' or ``equitable''
only with respect to {\em one particular moment}, however,
leaving out anyone born or reaching age of eligibility
after this one-time money creation event.
As a result, it seems clear that creation of democratic money
should be a continuous or periodic process.
In traditional elections, people are generally eligible to vote
periodically throughout their lifetimes after reaching some voting age.
A democratic currency should similarly offer
regular benefits to all members of a relevant population
throughout their lifetimes.
Thus, as with elections,
money creation in a democratic currency should probably occur periodically.
In principle, this process could even occur
at the same time and with the same
registration mechanisms, eligibility criteria, etc.
Whether such a close alignment with elections is actually desirable
is a more complex question we do not attempt to address at the moment.

As with the right to vote,
we expect the currency-distribution benefits of democratic money
to terminate upon the person's death.
Just as it is considered unacceptable to cast a vote for a dead relative,
it is equally unacceptable to draw on a dead relative's share of basic income:
the survivors must make their way 
on the basis of their own incomes, basic or otherwise.

Important differences between the properties we would want
from democratic money versus voting in elections
are greater fungibility and
at least moderately longer-lived utility as a store of value.
In particular, as mentioned above, unlike votes in elections,
it should be both allowed and expected for people
to trade their democratic money for arbitrary goods and services.
Further, democratic money should be fungible,
in that people should not have to know or care
{\em which} periodic money-printing ``batch'' a given coin came from:
democratic money should be more-or-less the same whether a participant
received it in this year's distribution event, last year's,
or indirectly from someone else through a trade or investment.
Finally, democratic money should be useful as a {\em real} store of value --
not just for an ephemeral duration as in the time period
between when votes are cast and an election's outcome is announced --
but at least long enough to support normal commerce and investment.
How much and how long-lived this store of value should be
gets into questions of fairness and equality over time
that we address in the next section --
but at the very least, it seems clear that democratic money
should offer significantly more long-lived value-storing capacity
than the ephemeral votes cast in a traditional election.

\subsection{The Ancient Principle of Equal Opportunity over Generations}

In addition to the principle of equality over population,
democracy also typically embodies an often-implicit principle
of {\em equality over time}:
namely, that voters in one periodic election should have
essentially the same power of choice as those
in prior or subsequent elections --
or as prior or subsequent generations of voters.
The policies one elected government institutes,
the next elected government can change or reverse,
according to the evolving will of the people.

Envision, to the contrary, a democratic system of government
in which newly-elected governments could only at most {\em add to} or refine
existing laws, but had no power to repeal them;
could add new spending items to the budget but not cancel prior ones;
could impose new taxes and/or new tax exemptions
but never clean the slate and simplify the tax code.
The policies of successive governments would effectively be squeezed into
a tighter and tigheter policy space,
its democratic decision-making power gradually crushed under the weight
of the decisions, good and bad, made by the representatives of past generations.
Such a ``democratic'' system would fail to ensure equality over time,
by effectively allowing the ghosts of past generations
to dominate and subjugate the democratic power of this generation.
In fact, it has been pointed out that some democratic processes,
such as popular initiatives that overly constrain the elected legislature,
can have precisely this deleterious effect over time,
rendering democratic government progressively more gridlocked
and dysfunctional~\cite{schrag98paradise,citrin09proposition}.

However, such a tendency for past generations' successes and mistakes
to progressively dominate future generations' opportunities and effective power
is a central tendency of modern capitalist economics, 
as has been observed by innumerable scholars
and definitively quantified
in the work of Thomas Piketty~\cite{piketty17capital}.
The winners of past generations find ways to use their existing advantage
to protect and increase their share of wealth,
conveying these advantages to their lucky heirs,
while leaving the children of the less fortunate to fight over
ever-smaller pie-slice of global wealth.
In short, ``the rich get richer.''
Notwithstanding the (substantial) flaws of current democratic structures,
today's economics unquestionably fails to ensure that successive generations
have the same economic opportunities as their predecessors,
and often fails even to offer a pretense
that equality of opportunity over time and generations
should be a central economic principle.

This was by no means always the case:
economic renewal through periodic cancellation of debt
(and freedom from debt-induced slavery)
was a basic and pervasive economic principle
for thousands of years in antiquity~\cite{hudson93lost}.
Sumerian, Assyrian, and Babylonian rules throughout Mesopotamia,
including the famous Hammurabi,
regularly proclaimed acts of general manumission,
cancelling all consumer debts and freeing debt-slaves,
in part to protect their own despotic power
from encroachment by internal economic aristocracies,
and in part as a military tactic to ensure a ready supply of peasant infantry
and limit the tendency of debt-slaves to defect to the armies
of neighboring agressors.

\begin{quote}
{\em Misharum} acts released cultivators
from the threat of debt-servitude resulting from financial arrears.
This gave them a stake in the society whose boundaries
they were fighting to extend.~\cite[p.21]{hudson93lost}
\end{quote}

After the power of central palaces gradually gave way to economic aristocracy
in this region around the first century BC,
Jewish reformers such as Nehemiah and Ezra
embraced this Babylonian tradition of periodic economic renewal
in the written law that they were codifying in largely its current form.
Beyond being a miltary tool for palace rulers to employ at their discretion,
however,
these Jewish reformers extended the principle of periodic economic renewal
to a populist religious covenant in the form of the
periodic ``year of jubilee'' --
a fixed 49- or 50-year cycle of freedom from debt and debt-servitude --
which in principle was to be an inviolate law of God
that not even a ruler had the authority to neglect.
Jesus, in his sole act of physical violence
recorded in all four Gospels,
drove the bankers and money-changers out of the temple of Jerusalem
as an act of cleansing,
in attempt to reconfirm and reestablish
the cultural and religious tradition that people had a basic right
to protection from unrestricted economic domination by the wealthy.
In short, the tradition of periodic economic renewal was a fixture
of Babylonian, Jewish, and early Christian culture
lost only in modern reinterpretations of those traditions.

\begin{quote}
Few Christians today recognize that when they pronounce the word ``Hallelujah,''
they are repeating the ritual term {\em alulu}
chanted to signify the freeing of Babylonian debt-slaves,
a rite followed by anointing the manumitted
individual's head with oil.~\cite[p.30]{hudson93lost}
\end{quote}

Of course, the Babylonians' tool of economic renewal
through periodic debt-cancellations would be disruptive
to the smooth functioning of modern economies, to say the least.
It would undoubtably be difficult for anyone to obtain a home loan
shortly prior to a (scheduled or anticipated) act of {\em misharum},
for example.
Periodic debt-cancellations or jubilee years
would also ensure fairness and equal opportunity
only over the long term, across generations, and not over shorter periods:
a person who falls into extreme debt shortly before a jubilee year
would obviously be much luckier than a person who does so just after one.
While in some sense economic disruption -- for the purpose of renewal --
is in part precisely the point,
and some short-term unfairness may well be tolerable
in the interest of long-term stability,
nevertheless there are probably alternative ways to achieve periodic renewal --
and protect a principle of equality over time and generations --
that do not inherently require a global economic ``hard reset'' every 50 years.

\subsection{Achieving Equality Over Time with Leprechaun Gold}

These ancient principles lead us to the obvious question
of whether the principle of equality over time and economic renewal
can be embedded into a modern currency
embodying the useful properties we generally expect money to have.
We will focus for now purely on monetary wealth,
leaving other forms of property such as goods or real estate
to address later in Section~\ref{sec:property}.

Recognizing that periodic economic resets such as jubilee years
would be difficult to reconcile with modern economics in numerous respects,
we first examine a straw-man alternative,
which we might describe as {\em leprechaun gold}:
a hypothetical form of currency in which each individual ``coin'' created
has a limited lifetime, after which it silently evaporates.
The fundamental idea here is that however money is
created, distributed, and utilized subsequently,
it should confer its holder a {\em temporary} share of economic power
in apportioning goods and services and prioritizing society's endeavors.
That is, such a leprechaun coin should confer
a share of economic power that lasts long enough to allow any individual
to reap rewards from prudent behaviors and wise investments
within his or her lifetime,
but should {\em not} translate into a share of aristocratic power
that can be passed down across unlimited successive generations.

On this basis,
if we were to pick a particular lifetime for such leprechaun gold,
then the Biblical jubilee period of about 50 years
might be a quite reasonable choice,
in that it happens to correspond roughly to the working lifetime
of a (modern, not Biblical-era) person.
\com{
In the \coin reference design we will actually use
a shorter 20-year nominal lifespan for circulating currency
for reasons discussed later in Section~\ref{sec:money-accelerated},
and apply the 50-year lifespan to {\em property} wealth
in Section~\ref{sec:property}.
}
There is nothing magic about these specific numbers, however,
and we can probably expect a fairly wide range of values to work fine
as long as they are globally fixed and within reason.

\begin{figure*}[t]
\centerline{\includegraphics[width=0.75\textwidth]{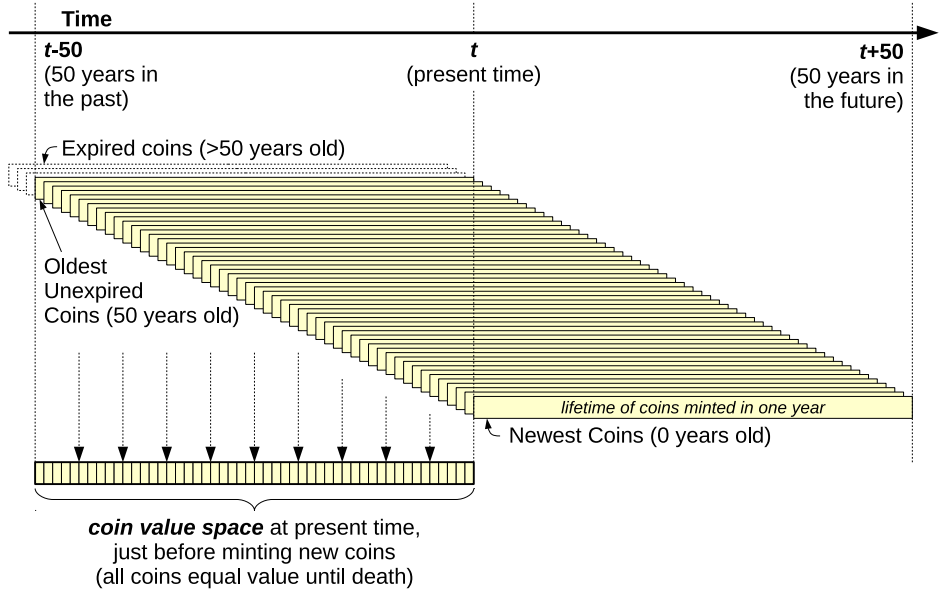}}
\caption{Illustration of the lifetimes and value space of ``Leprechaun Coin''}
\label{fig:leprecoin-time}
\end{figure*}

Combining the leprechaun gold idea
with the principle of equality over population
from Section~\ref{sec:eq-pop},
a democratic currency based on leprechaun gold
would be created periodically in batches
(\eg, once a year or once a month)
and handed out in equal measure to all individual participants
alive and eligible at that time,
then remain usable and hold their value for the coin's lifetime.
Each year, a portion of the previously-existing coins --
namely those that have exhausted their lifetimes --
would suddenly vanish or otherwise lose all value,
effectively ``making room'' in the economic value space
for the latest batch of newly-minted coins.
At any given moment,
the total economic value space would consist of
a ``window'' of fifty one-year batches of coins,
as illustrated in Figure~\ref{fig:leprecoin-time}.
This approach would by design guarantee (monetary) equality over time
by ensuring that the money newly-minted and distributed each year
always represents a fixed, one-fiftieth share
of the total economic value space represented by all extant currency,
while allowing for overall economic continuity
without requiring synchronized global hard-resets.

With conventional monetary technologies such as minted coins and banknotes,
one obvious problem with the ``leprechaun gold'' approach to economic renewal
would be simple technical impracticality.
It is unlikely to be easy or cheap to manufacture a coin or banknote
that literally vanishes at any time,
let alone after a precise time period
not readily manipulable by the holder(s) of the currency.
Coins could simply be engraved with an expiration date, of course,
but people would no doubt find it inconvenient to have to
read and check the expiration date of every coin or banknote they are handed.
However, these technical impracticality issues vanish in the post-Bitcoin era:
it would be readily feasible, and in fact technically quite trivial,
to create a cryptocurrency similar to Bitcoin
except with a built-in set of rules making coins become unusable
after a globally-agreed-upon time period.

The more fundamental problem with leprechaun gold
is that it would compromise the fungibility of the currency.
Older coins with closer expiration dates would always have less effective value
than newer coins with longer to live.
Sellers of a good or service
would demand more coins of a given nominal denomination if they are older
and fewer if the coins are newer.
Everyone would effectively have to treat the fifty extant batches of currency
at any given moment as fifty separate currencies to be juggled,
an obviously impractical task from a usability perspective.


\subsection{Democratic Monetary Policy: Equality over Time via Currency Devaluation}

While leprechaun gold is thus clearly no more practical for a modern currency
than reinstating Biblical jubilee years,
it does suggests a simple principle
for implementing equality over time in a democratic currency:
namely, that the new money minted and evenly distributed in one year
should always represent a fixed (\eg, one-fiftieth) proportion
of the currency's total economic value space after the minting.
Leprechaun gold implements this principle
by making all coins suddenly vanish after a 50-year lifetime,
but this is by no means the only way to implement the principle.

Another simple approach,
which ensures economic rewewal while preserving the fungibility of the currency,
is simply to {\em compress} the previous economic value space each year
to make room for the newly-minted currency.
In this approach, old coins never vanish or lose value entirely,
but instead smoothly and gradually lose their value over time
with respect to the value of newly-minted coins.
Suppose we stipulate that the democratic currency should have
a 50-year nominal lifetime like the leprechaun gold above,
roughly matching the working lifespan of a modern human.
Then each year we could simply print and distribute an amount of new money
that sums to exactly one-fiftieth of the value
of all existing or potentially-existing coins in the prior value space.

\begin{figure*}[t]
\centerline{\includegraphics[width=0.75\textwidth]{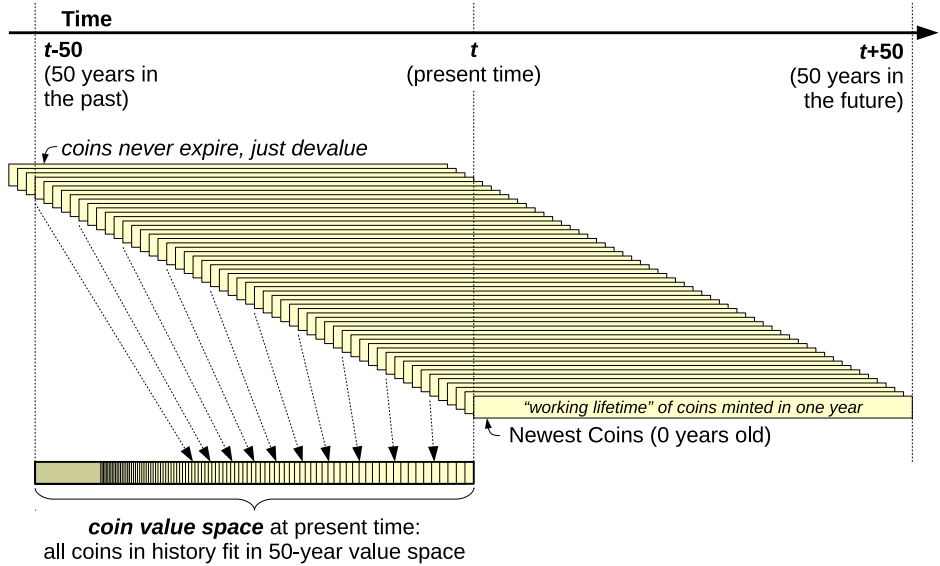}}
\caption{Illustration of value space of constant-inflation democratic currency}
\label{fig:lifecoin-time}
\end{figure*}

In modern economic terms,
this simply amounts to imposing a fixed yearly rate of monetary devaluation,
in this case 2\%,
resulting in a corresponding rate of inflation of the prices
of anything purchased in the currency.
While other factors such as the size of the population using the currency
and changes in the way they use it
will also in practice affect the currency's value and the prices of goods,
if all other such factors were held constant
the currency would experience a fixed 2\% inflation rate over time.
This built-in inflation rate may alternatively be viewed
as a periodic tax on the value of all existing coins in the currency
to pay for the regular minting and distribution of fresh currency.

It is interesting that in this currency design experiment,
we seem to have arrived at a target inflation rate
precisely matching the consensus that central bankers and economists
have arrived at in recent decades for conventional economies --
though in pursuit of a different goal,
namely the pragmatic objective of ensuring that economies
have enough monetary ``fuel''
to avoid negative interest rates and deflationary spirals,
not not so much as to cause hyperinflation
and bank runs~\cite{federal15why,noyer16thoughts}.
We might speculate as to whether central banks
have effectively ``discovered'' through practical experience
a rule-of-thumb target inflation rate that in fact works well
for some more fundamental underlying reason:
such as (for example)
because a 2\% inflation rate effectively causes money to retain its value
over a nominal lifetime roughly matching a modern person's working lifespan,
giving individuals the monetary tools
to invest and reap rewards in their lifetimes
while ensuring that subsequent generations have access to enough money
to enjoy similar economic opportunities.
On the other hand, it is unlikely that there is anything magic
about the value 2\% in particular,
and strong arguments have been made for higher inflation targets,
such as 4\%~\cite{ball14case}.

\com{

There's nothing magic about 2\%, of course.
There are arguments that, say, doubling the target to 4\%
would offer central banks more leeway and avoid the ``zero bound'' problem
without significantly harming an economy.

Converted to an intergenerational fairness benchmark,
this corresponds to money having a ``working lifetime'' of 25 years,
a bit shorter than but in the ballpark of the working career lifetime
of a modern person between the end of advanced education (\eg, 35)
and retirement (\eg, 65).
This would naturally shorten the time horizon with which one could
expect to ``make a fortune'' through successful entrepreneurship
and then enjoy it maximally post-retirement,
but may arguably be fine if the currency's basic income alone
eventually becomes sufficient to sustain individuals post-retirement
and the currency's main goal is to enable and fuel effective reinvestment
{\em during} an individual's working career.
There are reasonable arguments to be made both ways.

}

\xxx{ discuss the markdown approach as well }

\xxx{ reiterate that we're only talking about currency here,
	not other forms of property, and the limitations that entails,
	to be addressed later...}

\subsection{Value and Supply Stability in Democratic Currencies}

In a democratic currency operated as described above,
at any given moment there is a limited supply of currency,
although its total supply obviously increases over time
with each successive distribution of new coins to the population.
Provided at least some participants
also find the currency useful in some way,
for buying and selling certain goods and services for example,
there will be some level of demand for the currency.
Because participants must hold some amount of the currency in order to use it
and its supply is limited at any point in time,
the currency will have some effective {\em value} at any time,
determined by the amount of physical goods (or money in other currencies)
participants are willing to trade one unit of the democratic coin for.
Provided there are at least some participants
who believe in the value of the currency and want to use it for something,
its effective value will be greater than zero,
though not necessarily large.

Because the supply of new democratic currency is fixed and readily predictable,
the demand, and hence value, of the currency will depend on its use.
If participants initially use the currency for one niche special purpose
(\eg, to buy and sell virtual curiosities like cryptokitties~\cite{mala18who}),
then its demand and hence collective value will initially be quite low.
However, if participants then find additional uses for the currency,
demand increases while the rate of supply remains fixed,
thereby increasing the scarcity and effective value of the currency,
as measured in goods or other currencies for example.
Thus, while we have reason to believe a democratic currency formulated this way
would eventually offer considerable long-term stability,
we are expressly {\em not} setting price stability as the primary end-goal,
but are instead allowing the value of circulating currency to vary with,
and effectively be defined by, the collective use the population makes of it.

Because the supply of the democratic currency grows at a stable rate,
human population does not change that quickly,
and the ways and amounts in which the population uses the currency
is likely to be moderately similar and ``average out'' over the population
once the currency is widely deployed,
there is reason to believe that the currency's value may be moderately stable
over the long term
(though, once again, not necessarily large).
In traditional fiat currencies,
money supply is controlled indirectly by central banks
through money-printing and the interest rate of central bank loans,
but its effective supply is ultimately dependent on the economic sentiments
of a relatively small, elite group of bankers making decisions
on loans and investments.

A democratic currency effectively short-circuits these financial intermediaries
by distributing new currency directly and evenly to the entire population,
thereby inherently spreading the basis for its value and stability
across a much larger and by definition more decentralized population.
While one or a few individuals may change their currency-using behavior
drastically over a short period,
it is much less likely or common that huge populations do,
except in response to a severe shock of some kind --
and such traditional economic shocks usually propagate {\em from}
the financial world to the larger population, not vice versa.
The democratic currency also by design ensures that
no one needs to worry about their supply of currency drying up entirely
since everyone always has a constant supply of basic income,
whatever it may be worth at present in terms of purchasing power.


\subsection{Relationship to Stamp Scrip and Accelerated Money}
\label{sec:money-accelerated}

The idea of regularly devaluing circulating currency is hardly new.
In the early 1900s,
Silvio Gesell proposed a monetary system
he called ``free money'' ({\em Freigeld}),
in which the holders of banknotes must buy and affix stamps
to banknotes each week
to maintain their validity over time~\cite{gesell58natural}.
Gesell's purpose for this ``carrying tax''
was to incentivize the productive use of money
as a medium of exchange in commerce
and penalize hoarding --
effectively by making banknotes like ``hot potatoes''
that the current holder wants to pass on as quickly as possible.
Because loans in this hot-potato currency
would transfer the carrying tax to the receiver of the loan for its duration,
holders of excess money would be incentivized to loan it interest-free
to a worthy cause instead of holding it themselves.

In several small-scale experiments where it was implemented as intended,
such as Schwanenkirchen and W\"orgl,
these {\em stamp scrip} currencies indeed seemed
to accelerate local economies almost miraculously,
until in most cases the experiment was squashed
by central monetary authorities~\cite{champ08stamp}.
Subsequent stamp script attempts in the US were less successful,
in part because they incorrectly used stamps to implement
a {\em transaction tax} rather than a {\em holding tax},
incentivizing the exact opposite behavior as Gesell's design.
At least one community currency based on Gesell's ideas
has stood the test of time, however:
WIR Bank was founded in Basel in 1934
to create an accelerated community currency complementary to the Swiss franc,
and remains in operation to the present,
although it has evolved into more of a credit network for interest-free barter
without the stamp scrip mechanism~\cite{defila94wir,studer98wir}.

Gesell also strongly influenced John Maynard Keynes,
whose macroeconomic theory adopted the explicit devaluation idea
for the same purposes of ensuring circulation and avoiding ``liquidity traps,''
though using the mechanism of deliberate monetary inflation
instead of stamped banknotes~\cite{keynes65general}.
In this way, Gesell's ideas have indirectly entered
the basic {\em modus operandi}
of inflation-targeting central banks~\cite{federal15why,noyer16thoughts}.

The constant, explicit devaluation we propose for democratic currency
clearly bears close relationship to the Gesell and Keynes tradition,
and we expect its adoption could have a similar
acceleration of commerce effect on the use of democratic currency.
However, in our case neither the acceleration of commerce,
nor the stabilization of purchasing power that Gesell and Keynes also sought,
are the primary motivating end-goals, but only desirable side-effects.
In democratic money, the essential purpose of fixed-rate devaluation
is to provide a systemic guarantee of equal opportunity over time,
by ensuring that the fresh democratic currency distributions each year
always represent an equal-size ``slice'' of the total monetary pie
at that time,
and that the economic successes and failures of past generations
always make room for, rather than gradually strangling,
the economic opportunities of their successors.

Since we intend to implement a democratic currency
initially as a cryptocurrency,
it is worth contrasting with Bitcoin~\cite{nakamoto08bitcoin},
which is hard-coded with a strongly {\em deflationist} monetary policy.
Bitcoin exponentially decreases its fresh money supply over time
so that no more than 21 million Bitcoin will ever be created.
Since Bitcoin can and regularly is lost or destroyed in various ways,
the amount of circulating Bitcoin can only decrease in the long run,
which has strongly incentivized its use mainly for speculative investment
or ``HODLing'' (Holding On for Dear Life) in the cryptocurrency culture,
eventually to the near-complete exclusion
of productive uses as a medium of exchange.
As a result, Bitcoin has become more comparable
to a collector's item than a currency:
a curiosity not unlike a painting or a cryptokitty~\cite{mala18who},
whose price rests entirely in its scarcity
and what collectors are willing to pay for it.
As finance expert Alexander Lipton observed,
``Bitcoin has no value, hence it can have any price!''~\cite{lipton18banking} 
In contrast, we want a democratic currency to be a {\em working} currency
whose value and price support is primarily in the commercial utility
it is providing the people using it.
Fixed-rate devaluation in the tradition of Gesell and Keynes
support that purpose together with our primary ultimate goal
of providing long-term assurance of equal opportunity over time.



\xxx{ Discuss Lietaer and incentivizing sustainable investment. }

\subsection{Relationship to Universal Basic Income}

\xxx{ see and perhaps cite Tim Jackson's article "Confronting inequality: basic income and the right to work" }

The idea of {\em universal basic income} or {\em UBI}
has recently been attracting widespread interest,
as a way to simplify the governmental tax and redistribution mechanisms
embodying a social ``safety net.''
The idea of UBI is for the government to give each citizen
a flat basic monthly income,
with no regard for how much they need it or how they use it.
Part of UBI's appeal comes in the form of
eliminating the complex governmental mechanisms for evaluating need,
and closing the loopholes enabling abusers to extract undeserved rewards
by manipulating the system and claiming need falsely.
Another appeal is that UBI eliminates
the disincentives to productive work
that current safety net schemes often accidentally embody:
\eg, by disqualifying a person for unemployment benefits
if they hold {\em any} job,
and hence penalizing them for taking a job whose monthly income
would be less than what they obtain through unemployment.
Since a UBI is not affected by employment status,
a currently-unemployed person would always see their total income increase,
and never decrease,
by taking on any job regardless of salary.
Small-scale UBI experiments are underway in Finland~\cite{koistinen14good,donnelly18finland}
and California~\cite{ycr17basic},
and an initiative for large-scale implementation
was recently proposed via initiative
in Switzerland but defeated at the polls~\cite{bbc16switzerland,wagner16swiss}.

As with many governmental tax-based redistribution schemes,
two of the most problematic questions surrounding UBI
are {\em how much should the benefits be?}
and {\em where should the money come from?}
Plausible answers to the latter question typically require eventually cancelling
a variety of existing social safety-net programs
and redirecting their tax revenue to the UBI:
otherwise, the claimed simplification benefits of UBI
obviously would not materialize.
However, cancelling even one existing safety-net program, let alone many,
is fraught with enormous practical risks and political quagmires,
starting with the question of how effectively a UBI would actually substitute
for existing safety-net programs it is meant to replace.

\subsubsection{How to Decide the Amount of Basic Income?}

A perhaps even more fundamental challenge with UBI, however --
and a problem that would not disappear
even once the practical and political transition hurdles are surmounted --
is answer the question of {\em how much} the UBI should be.
Too little, and the UBI will inadequately serve its safety-net function;
too much, and it could impose an unsustainable cost on the economy
or incentivize too many people to stop seeking traditional work entirely.

And by what principle should the UBI be adjusted over time?
Leaving it at a fixed face-value amount in a traditional currency
will allow inflation to erode the UBI's
purchasing power and effectiveness with inflation over time.
Keeping the UBI inflation-adjusted might solve that one problem,
but would fail to account for other types of social and economic evolution.
For example, its practical effectiveness
could still gradually lose synchronization
with society's quality-of-life expectations,
and with the abstract basket of goods and services
that people in practice consider ``basic'' and ``essential.''
For example, Internet access was until recently considered a luxury
but now is generally considered essential,
and the ad-driven evolution of the Internet
ecosystem increasingly makes many Web sites and applications
effectively unusable to anyone without sufficient bandwidth
or a powerful smartphone~\cite{rogers18what}.

Perhaps worst of all, an attempt to define and maintain through policy
an explicit basket of goods and services from which to define the UBI's value
could a policy free-for-all
as every special interest whose product isn't
in the basket of ``essentials'' tries to legislate it there
(the equestrian society claiming that
every child has a basic right to a pony, etc.),
as every special interest whose product {\em is} in the basket
inflates the prices of its products as the US medical industry has done
(because, after all, {\em someone else} --
\ie, the public -- is paying for them via UBI),
and as the slowly-inflating UBI basket of products with slowly-inflating prices
gradually consumes the economy like a slow-boiled frog.

A key advantage of the proposed democratic currency approach
is that no policy decision is needed to set, or maintain, the level of UBI,
in terms of either face-value, purchasing power, or lifestyle expectations.
The democratic currency has a {\em floating universal basic income},
defined indirectly as a fixed percentage of the total value
the currency is collectively providing those who are using it.
If the currency is being used by only a few people for little of importance,
as will inevitably be the case initially if launched as a cryptocurrency,
there will at first be little demand for it aside from speculation,
and the value of the floating basic income
denominated in goods or other currencies may well be extremely small.
That is fine, as long as {\em a few} people find {\em some} utility in it
and we avoid setting unrealistic expectations.
As people find more uses for it, demand for the currency grows
while supply remains a steady flow in terms of person-hour-coins per year.
The value of each person-hour-coin thus increases in purchasing power,
and the effective value of each person's floating UBI grows correspondingly.
The democratic currency's floating UBI promises
neither a free lunch nor any particular standard of living,
but instead defines a UBI that a society can {\em sustainably afford} --
as precisely the fixed slice of collective value-space
by which the currency is devalued each year to feed economic renewal
in ensuring equal opportunity over time.

The size of that fixed slice, in turn,
is calibrated to a fundamental property of humans
that so far remains extremely stable and slowly-changing by economic timescales:
namely, the typical ``working lifespan'' of a person,
and the respective ``nominal lifetime'' of coins distributed to people,
which we take to be 50 years in the reference design.
The correctness of this calibration metric could of course be disrupted
if anti-aging science comes to fruition
and people start living hundreds of years --
but let us worry about crossing that bridge once it is in sight.

\subsubsection{Cross-Border Effects and Migration Incentives}

A further hazard arising from implementing a basic income at government level
is that it will inevitably not be truly ``universal,''
but at best universal only within the jurisdiction
of the country, state, or city that (first) implements it.
Even when neighboring governmental jurisdictions both enact a UBI,
the policy-determined amount of the UBI will almost certainly be different.
Richer countries or cities will want to set their UBI high enough
to satisfy the high standard-of-living expectations of their citizens,
while their poorer neighbors
will have to make do with a lower UBI at best,
due to being unable to afford a higher one if nothing else.

These variances in the existence and amounts of UBI
are likely to create strong migration incentives
toward districts with a (larger) UBI,
which are after all handing out money to anyone who can ``prove'' residency.
One likely silver-lining benefit of the sheer complexity
of current safety-net programs provided by governments,
and the many frustrating administrative hoops one must often jump through
to demonstrate need and take advantage of them,
is that not {\em too} many people are probably tempted to pick up
and move to the next city, state, or country
just to free-load on more-generous and well-funded safety-net programs there
(though no doubt some do, such as Americans who travel or move abroad
in part for more affordable healthcare
or other safety-net services~\cite{braverman16more,schwab17these}).
But once the value of one jurisdiction's safety-net programs
has been transformed into
completely liquid, no-questions-asked handouts of unrestricted cash,
the benefit from being a resident of a high-UBI district --
or even ``proving'' residency in several UBI districts at once
by obtaining fake IDs or other trickery --
will be much more clear and compelling
to anyone even a bit ``morally flexible.''
The citizens of the high-UBI jurisdictions
will naturally perceive this influx of invading UBI-seekers negatively,
fueling even more anti-immigrant sentiment and fortress mentality
than we already have.

By implementing a floating basic income
in a democratic currency as described here, in contrast,
the UBI by definition be the same anywhere in the world the currency is used,
would be associated with the currency and not any given jurisdiction,
and would be defined by the currency's fixed devaluation-and-renewal rate
instead of varying policies tuned to reflect varying expectations.
People and communities anywhere in the world would be free to adopt it
as much or as little as they see fit
(and as their respective governments permit them).
Those that do adopt and use it contribute incrementally
to demand for the currency,
and hence to its total value as denominated in goods or other currencies,
thereby incrementally increasing the value of the floating basic income --
for {\em everyone} using it at once.

The migration incentives
resulting from a democratic currency's floating basic income
will likely, in fact, flow in the opposite direction as for conventional UBI:
namely toward {\em poorer} jurisdictions rather than richer ones.
This is because while the currency will provide the same UBI for all users,
as denominated in other currencies tradeable with it,
the effective {\em purchasing power} of that basic income will be higher
anywhere the cost of living is lower.
Thus, anyone substantially relying on basic income from \coin
will be incentivized to to move to poorer regions or countries
where that income wields more power to purchase the necessities of life.
If widely adopted, a democratic currency could thus help counteract
and perhaps even reverse some of the migration pressures
that are currently stoking anti-immigrant and strong-border attitudes globally.

\xxx{ Discuss: why invest?  (a) growth of participation, and
		(b) economic value due to growing use }

\subsubsection{Cold Hard Numbers: Guesstimating Basic Income from Current Economic Statistics}

Accepting that \coin's basic income
will not and is not intended to
guarantee any particular purchasing power or quality of life,
can we estimate what this basic income {\em might} amount to
if the currency were to become widely-adopted in some context?
Performing a detailed and rigorous economic analysis or forecast
would be nontrivial to say the least
and is outside both the scope of this paper
and the domain of this author's expertise.
However, we can use readily-available data to produce a few ballpark estimates
under admittedly loose and
perhaps wildly optimistic ``thought-experiment'' assumptions.

Suppose hypothetically, for example,
that \coin were in widespread use {\em today},
as the primary money supply in circulation,
either globally or within particular countries of interest.
We make the simplifying assumption that this widespread use of \coin
does not significantly change the size or behavior of the economy
{\em other than} by producing a basic income fed by fixed-rate devaluation.
This simplifying assumption is of course unrealistic,
as adopting a significantly different monetary system
would obviously have many far-reaching and difficult-to-predict impacts.
We hope and expect that the balance of \coin's impacts would be positive,
\eg, making the economy larger and more active due to
the monetary acceleration effect and the fact that
everyone would have a minimal purchasing power base matched with need to spend
on the essentials of life.
But other factors would of course cause effects in the other direction:
\eg, currency devaluation higher than current interest rates
could incentivize transfer of value away from money to other forms of wealth,
thus reducing the amount of global wealth held in circulating money,
especially if provisions for the incorporation and gradual devaluation
of property-wealth are not also made as described in Section~\ref{sec:property}.
However, acknowledging the large uncertainties involved,
let us pretend that these effects did not exist
and that today's global economy {\em was} based on \coin.

\begin{table*}[t]
\begin{center}
\begin{tabular}{l|r|r|rrr|r}
		&	 	& 		& \multicolumn{3}{|c|}{Basic Income}		& \multicolumn{1}{|c}{Official} \\
		& M1 Money	& Population	& \multicolumn{3}{|c|}{(USD/Year)}		& \multicolumn{1}{|c}{Poverty Line} \\
Region		& (USD Million)	& (Thousands)	& 2\% rate	& 5\% rate	& 10\% rate	& \multicolumn{1}{|c}{(USD/Year)} \\
		
\hline
\hline
Global		& 36,800,000	& 7,630,000	&    96.50	&   241.00	&   482.00	&    694.00 \\
\hline
Switzerland	&    657,000	&     8,540	& 1,540.00	& 3,850.00	& 7,690.00	& 26,900.00 \\
United States	&  3,660,000	&   327,000	&   223.00	&   560.00	& 1,120.00	& 11,800.00 \\
India		&    440,000	& 1,350,000	&     6.51	&    16.30	&    32.60	&    172.00 \\ 
Nigeria		&     30,200	&   196,000	&     3.08	&     7.70	&    15.40	&	 \\
\hline
\end{tabular}
\end{center}
\caption{Ballpark basic income estimates based on M1 money in today's global economy}
\label{tbl:basic-income-est}
\end{table*}

Table~\ref{tbl:basic-income-est} shows ballpark estimates
of what a \coin basic income might look like, denominated in today's US dollars,
if \coin were aready in use globally or in a somewhat-arbitrary selection of countries.\footnote{
	\xxx{ Data sources:
		\url{https://tradingeconomics.com/country-list/money-supply-m1},
		\url{http://www.visualcapitalist.com/worlds-money-markets-one-visualization-2017/},
		\url{http://www.worldometers.info/world-population/population-by-country/}}}
We use recent statistics for M1 ``narrow'' money supply,
which includes immediately-accessible money such as banknotes, coins, and checking accounts,
but excludes less-liquid forms of money such as savings accounts.
This choice reflects the intended implementation of \coin as a cryptocurrency similar to Bitcoin,
in which banks, exchanges, or electronic wallets can neither create nor destroy currency
but can only {\em hold} it on behalf of users.
Like Bitcoin, 
\coin would thus behave as a full-reserve currency~\cite{douglas39program,benes12chicago},
in which the contents of immediate-access checking accounts or digital wallets
would be directly subject to the currency devaluation and democratic distribution processes,
but less-liquid, loan-based accounts would not be directly affected.
Based on these statistics,
we estimate what \coin-derived basic income would amount to yearly in USD,
for three different comparative rates of devaluation and democratic redistribution:
a conservative 2\% rate representing a nominal currency lifetime of 50 years,
a more aggressively-accelerated 5\% rate closer to the suggestion of Silvio Gesell~\cite{gesell58natural},
and an extremely-accelerated rate of 10\%
on the boundary of what economists would call ``galloping inflation.'' 

These estimates make it clear that we should make no pretense of expecting
a basic income derived solely from circulating \coin, even if ubiquitously adopted globally,
to offer a satisfying quality-of-life in developed countries
or to replace their existing safety-net programs.
The basic income \coin provides might be hardly noticeable financially
to the average resident of the developed world, in fact.
However, when compared to the UN's official global poverty line of \$1.90 per day or \$694 per year,
the basic income derived from a globally-deployed \coin could cover a significant fraction of the distance
toward eliminating global poverty by this definition, if not quite achieving it.
In fact, the \coin basic income at the Gesell ``Freigeld'' rate of 5\% 
would be sufficient to overcome the current official {\em national} poverty line of India,
which is well below the UN's global poverty line.
Viewed in this light, the basic income \coin could in principle
be meaningful to the finances and quality of life of a large percentage of the world's population,
if by no means a complete solution.

Of course, basic income level that might be expected from widely deploying \coin within a particular country
would of course vary widely based on how rich or poor the country is,
as illustrated by the estimates based on the M1 monetary supply and population of particular countries
in Table~\ref{tbl:basic-income-est}.
If ubiquitously adopted in Switzerland, one of the world's richest countries per capita,
\coin's basic income would be the envy of much of the developing world --
but would appear quite unimpressive {\em in Switzerland itself},
reaching only a small fraction of the way up toward Switzerland's national poverty line.
If a poorer country such as India or Nigeria adopted \coin internally,
the results might be similarly less than satisfying with respect to the global poverty line.
These estimates thus clearly underline the need for \coin to be ultimately deployed ``across boders'' --
among large populations in rich and poor countries alike --
before we can plausibly expect the basic income it furnishes to offer
financially significant purchasing power to a significant portion
of its population of users.
These numbers also underline the huge risks inherent in trying to implement conventional basic income ideas
at policy-defined amounts
corresponding to national quality-of-life wishlists,
both in terms of the internal affordability and sustainability of such programs,
and in terms of the intense migration pressures from poor to rich countries
that such programs would inevitably exacerbate.

Since widely deploying \coin across borders involving both rich and poor countries
would clearly create a significant wealth redistribution flow from the former to the latter,
why might we expect the residents of rich countries to be willing to adopt
an international democratic currency like \coin voluntarily?
Two main plausible reasons present themselves.
Potentially appealing to the more socially-conscious members of rich populations
is the simple argument that {\em it's the right thing to do},
in the interest of addressing global inequality problems and establishing a more sustantiable global economic foundation,
and would represent a readily-affordable cost to the residents of rich countries.
Potentially appealing to the more inwardly-focused
members of rich countries with strong anti-immigrant sentiments,
in contrast,
is the argument that \coin would help counteract the global migration pressure on the rich countries --
even incentivizing less-well-off people to move in the {\em opposite} direction
toward poorer countries where their \coin-derived basic income would be exactly the same but their cost of living much lower.
Thus, there seems at least hope that a basic income derived from a properly-implemented and widely-deployed \coin
could appeal in different ways to both the more left- and right-leaning populations alike in the richer countries.

\subsection{Implementing Democratic Money as a Cryptocurrency}
\label{sec:money-impl}

A government's central bank could in principle issue a new currency --
or even revise the monetary policy and distribution of some existing currency --
according to the above principles to create democratic money.
A government-initiated rewriting of monetary principles this fundamental
is unlikely to happen any time soon, however,
not only due to the momentum of conventional economic thought,
but also because such a transition would represent
a huge and risky transition in practice if taken suddenly.

The emergence of digital cryptocurrencies such as Bitcoin, however,
offer the opportunity for the relatively ``safe'' redesign and deployment
of new currencies embodying new principles at small scales.
Creating a new cryptocurrency requires only that
a fairly small group of people consider the idea for the currency interesting,
and invest their time and energy into its development and deployment.
A new cryptocurrency effectively creates a new metric of value
which can gracefully coexist with all the existing currencies
and quasi-currencies such as frequent flyer programs and such,
without needing the permission of any existing organization or government,
apart from conformance to the currently-emerging regulatulatory regimes
applying to cryptocurrencies.

The design of currently-deployed cryptocurrencies such as Bitcoin,
as well as emerging proposals designed to improve their
scalability~\cite{kokoris17omniledger,eyal16bitcoinng,kokoris16enhancing,gilad17algorand},
privacy~\cite{sasson2014zerocash,ruffing14coinshuffle,narula18zkledger},
and functionality~\cite{nikitin17chainiac},
should in most respects be sufficient to support the technical requirements
of a democratic currency as described here.
Bitcoin, for example, already has a particular fixed monetary policy
embedded in the software that all the miners and users run:
namely a strongly deflationary monetary policy,
where exponentially less new money is created by the miners over time
such that at most only about 21 million Bitcoin will ever exist.
Bitcoin's monetary policy obviously comes nowhere close to satisfying
our principles of equality over time and constant economic renewal,
and its deflationary model strongly incentivizes
speculative investment in it over any productive use as a working currency.
But the arbitrary monetary policy Bitcoin encodes in its software
can readily be changed to a different monetary policy
in a different, democratic cryptocurrency.

\subsubsection{Membership and Stakeholder Models for Cryptocurrencies}

A key technical challenge
to implementing a democratic cryptocurrency
is to enforce a {\em membership} or {\em stakeholder} model securely
in which each human participant can obtain one, but only one,
equal share of cryptocurrency minting or mining power.
The fundamental problem is that today's digital systems have no secure way
to distinguish between two (or a thousand) real people,
and two (or a thousand) fake identities held by only one real person.

Bitcoin attempted to solve this fundamental technology challenge
through {\em Proof-of-Work},
in which participants compete to solve cryptographic puzzles
in order to win temporary membership rewards and create new currency.
Bitcoin's Proof-of-Work model,
as well as most other proposed foundations for cryptocurrencies
such as Proof of Stake~\cite{kiayias16ouroboros,gilad17algorand},
Proof of Space~\cite{dziembowski15proofs,park18spacemint},
etc.,
are all {\em investment-proportional} membership and reward models:
anyone who can afford to invest more in the system
reaps at least proportionally greater rewards.
In fact, due to the economies of scale and the many competitive advantages
available to the biggest players
in current cryptocurrencies,
larger investors in cryptocurrencies can typically obtain
{\em disproportionately} larger shares in power and rewards,
which has rapidly re-centralized mining power and profits
into the hands of a few large
players~\cite{gervais14bitcoin,gencer18decentralization,vorick18state}.
In short, because their mining power and reward models 
are investment-proportional rather than population-proportional,
today's cryptocurrencies fail either to
``decentralize'' or ``democratize'' money reliably,
and instead merely reproduce the ``rich get richer'' principle
of unconstrained capitalism in the guise of new technology.

\subsubsection{Enforcing ``One Person One Vote'' via Proof-of-Personhood}

Thus, a democratic currency would actually have to address and solve
the fundamental technical challenge of distinguishing
real people from fake accounts.
One obvious way to do this would be for a government or other organization
to manage membership in the cryptocurrency
by checking government-issued IDs or taking biometric samples of participants
and checking them against a master database
to make it difficult for one person to obtain many accounts or memberships.
This approach, while plausible, embodies problems of security
(\eg, it is not in practice that difficult or expensive to obtain fake IDs),
centralization
(we must trust some authority to check biometrics and maintain the database),
and privacy
(the master ID or biometric database becomes
a prime target for hacking or leaking).

A potentially more decentralized and privacy-preserving alternative approach
is via {\em pseudonym parties}~\cite{ford08nyms},
or real-world events run by local or regional communities 
in which people can show up in person and obtain pseudonymous tokens.
Pseudonym parties 
enforce ``one person one vote'' equality over population
on the basis of physical security and the fact that
real people still, for the moment, have only one body
and therefore can be in only one place at any particular time.
The use and adaptation of pseudonym parties as a basis for cryptocurrencies
has already been proposed as ``Proof of Personhood''~\cite{borge17PoP}.
Development and experimentation with this model is ongoing,
but we leave the details of this important technical challenge
out of the scope of this paper.
Another alternative approach being explored,
similar in end goal but pursuing a fully-online mechanism,
is {\em Proof of Individuality}~\cite{nygren15proof},
in which participants verify each other's apparent humanness
through video conferencing.
It is unclear how long such ``online Turing tests''
can plausibly remain secure, however,
given that today's audio, visual, and artificial intelligence technologies
can already synthesize fake humans
that trick a real human into believing they're
conversing with another real human~\cite{carey18human}.
\xxx{ also cite fake videos, human faces, etc}

\subsubsection{Cryptocurrency Accounting in \coins and \atoms}

We have so far described \coin's economic principles
in terms of an abstract value space,
neglecting details such as how accounting is done in this value space.
We now descend slightly into these details.

We take Bitcoin as a reference cryptocurrency,
in which money never has any physical embodiment,
but purely takes the form of information
on a distributed ledger or {\em blockchain}.
As in Bitcoin, \coin defines rules coded into software implementations
for how and when money can be created, and how money may be transferred,
based on transactions entered on this digital ledger.
Both transactions that create money ({\em coinbase} transactions)
and those that merely transfer money (payment transactions),
when validated and successfully committed to the ledger,
create {\em unspent transaction outputs} or UTXOs
that subsequent transactions may spend by ``consuming'' as inputs.

\paragraph{Satoshis versus \atoms:}

Leaving operational details aside,
the important question for now is how to represent the {\em value}
of these transaction inputs and outputs in \coin ledger entries.
Bitcoin represents the value of inputs and outputs
as an integral number of {\em Satoshis},
the smallest atomic unit that can be transferred,
defined as a hundredth of a millionth ($10^{-8}$) of one Bitcoin.

Following Bitcoin's example, we find it convenient
to represent transaction values
in integral multiples of some atomic unit of value,
which we will call {\em \atoms}.
Like Satoshis, \atoms will be indivisible,
one \atom representing the smallest nonzero transfer that can be made,
enabling us to avoid the complexity of floating-point and
the management of roundoff errors in value accounting on the ledger.
We want \atoms to be small enough
to allow sufficient accounting precision for practical use
and subdivision as needed
even in the hypothetical ``worst-case'' scenario in which
the entire world population of 7.6 billion people
immediately adopted \coin as the single global currency and the unit of value
in which all currency and noncurrency wealth is measured.

Taking total global wealth
currently estimated at 280 trillion USD~\cite{creditsuisse17global},
we convert this into Iranian Rial (IRR) --
currently the weakest official currency globally at about 42,000
IRR per USD~\cite{fxssi18top} --
to arrive at about $1.2 \times 10^{19}$ IRR as current global wealth
measured in the most fine-grained modern currency unit
that anyone might plausibly wish to transfer.
Since this just below the convenient binary integer $2^{64}$
(about $1.8 \times 10^{19}$),
we define \coin's initial total value space
to be exactly $2^{64}$ \atoms in size --
enough precision to accommodate comfortably today's entire wealth economy,
denominated at IRR granularity in this value space,
in the fairly unlikely event that we should need to.

While Bitcoin's deflationary monetary policy limit its value space forever
to about 21 million Bitcoin or $2.1 \times 10^{15}$ Satoshis,
the size of \coin's value space grows gradually over time without bound.
This gradual expansion, of the number of \atoms representing the value space,
enables us to implement \coin's periodic monetary devaluation and redistribution
without ever having to update the \atom-denominated account balances
in accounts or unspent transaction outputs (UTXOs) on the \coin ledger.
At the 2\% devaluation rate corresponding to 50-year nominal currency lifespan,
this means that each year we multiply the total value space size in \atoms
by a factor of 50/49,
then create and distribute new currency amounting to 1/50th
of the new value space denominated in \atoms.
For example,
starting with a year 0 value space
of $2^{64} \approx 1.84 \times 10^{19}$ \atoms,
at year 1 the value space expands to
$2^{64} \times \frac{50}{49} \approx 1.88 \times 10^{19}$ \atoms,
at year 2 it expands to
$2^{64} \times (\frac{50}{49})^2 \approx 1.92 \times 10^{19}$ \atoms,
and so on.

\paragraph{Bitcoins versus \coins:}

Bitcoin defines a fixed ``exchange rate''
between its minimum-granularity unit of value (the Satoshi)
and its ``face value'' unit intended for user consumption
(one Bitcoin, or 100 million Satoshi).
If we similarly define one \coin to be a fixed number of \atoms,
then \coin will suffer the same mild but annoying
flaw in ``user experience'' (UX)
as all modern Keynesian inflationary currencies,
namely that the face-value of the currency is worth less over time.
All product prices, salaries, etc.,
must be periodically raised to account for this loss of face-value,
and economists must constantly inflation-adjust prices
to make useful comparisons of value across historical timelines.
While we have been living with
and could no doubt continue to live with this UX flaw,
the flexibility of cryptocurrencies --
in which the only face-value amounts users ever see
are computed dynamically by digital wallet and payment applications anyway --
gives us the opportunity to fix this flaw
without introducing other UX inconveniences
such as Gesell's stamped banknotes~\cite{gesell58natural}.

We therefore define one \coin as equal to a {\em time-varying},
rather than fixed,
number of \atoms.
In the tradition of time-based currencies,
we define one \coin to be equal to one day's supply
of one person's basic income,
whatever that may be worth in terms of purchasing power.
That is, at any given time,
one \coin is equal to the number of \atoms of money newly-created
in the most recent year's distribution
(\ie, 1/50 of the total current value space size),
divided by the number of participants in the most recent distribution,
divided by the average number of days per year (365.25).
This way, if and when use of \coin stabilizes in some user population,
in terms of both number of participants and their typical usage patterns,
one \coin in one year will represent similar economic purchasing power as
one \coin several (or many) years in the future,
requiring no inflation-adjustment of face value over time.

The minor downside of this time-varying definition of \coin
with respect to \atoms, of course,
is that the face-value balances of account balances denominated in \coin
will suddenly change at yearly devaluation and redistribution events,
as a result of the increase in total value space size
and any year-to-year changes in participating population.
Account balances will remain the same when counted in \atoms,
but will change year-to-year when denominated in \coin.
People are already accustomed to their electronic bank account balances
changing more-or-less automatically from month to month and year to year,
due to charging of fees, depositing of interest, etc.,
so such occasional face-value changes in account balances seem workable.
A question for further study is whether it would be beneficial,
or problematic,
to ``smooth'' these account balance changes
by adjusting the \coin-to-\atom ratio gradually throughout a year
rather than in one more-noticeable step each year.

\subsection{Population and Participation Changes and Adoption Incentives}

So far we have assumed that the number of participants using \coin
is relatively stable and fixed,
but of course this is not the case in practice.
In the long term, once a democratic currency is widely adopted,
its population of users might eventually be relatively stable and slow-changing,
since human population increases at a limited and relatively slow rate
in comparison with economic timescales,
and rarely decreases rapidly except due to catastrophic wars or disasters.
However, if democratic currency is introduced as a cryptocurrency
operating alongside traditional currencies,
as seems most practically feasible and experimentally safe,
then the currency's user population will start out tiny
and grow as people learn about and start using it.

In defining \coin's distribution and devaluation mechanisms,
we have made no special provisions for changing population,
other than requiring that each distribution of new currency
be divided equally among the population participating {\em at that time}.
To analyze the effect of this design,
first consider the long-term situation when the currency is widely adopted
and the user population is changing slowly due to human births and deaths,
and also that the {\em ways} in which the population is using the currency --
the types of goods and services they are selling and buying with it --
is also relatively stable.
Suppose that in some year $Y$, a population of one million uses \coin,
and that twenty years later at $Y+20$,
the population has doubled to two million.
At years $Y$ and $Y+20$ the same fraction (1/50)
of the currency's total value space is reclaimed through devaluation
and distributed equally,
among 1M people at $Y$ and among 2M at $Y+20$.
Thus, each person's basic income slice at $Y+20$
will be half the size,
measured in percentage of the total currency value space,
as each person's basic income back in year $Y$.

Does this mean each person's basic income at $Y+20$
will have twice the {\em value} as at $Y$?
Definitely not,
because the overall economy using \coin has also grown in the meantime.
Again assuming the people at year $Y$ and $Y+20$ use \coin 
in roughly the same amount and fashion,
at year $Y+20$ there will be twice the number of people
needing to buy and sell about the same amount of food, services, etc.,
thereby doubling the effective {\em demand} for \coin,
and hence doubling the total commercial value
that \coin's value space represents
as measured in products, services,
or another currency with stable value.
Thus, if population changes gradually while other factors remain fixed,
each person's basic income may be expected to have
about the same purchasing power at year $Y$ and at $Y+20$,
despite each person's slice of the yearly pie having been cut in half.
And because of the currency's regular devaluation,
people at year $Y$ will probably have long ago spent or invested
their incomes (basic or other) by the time $Y+20$ rolls around.

Consider now, however, what happens when user population changes rapidly,
\eg, as a cryptocurrency originates in a small group and ``catches on''
as more people discover and voluntary adopt it.
Suppose, for example, the user population doubles in just one year,
from $Y$ to $Y+1$.
In this case, the basic income pie-slices at year $Y$
are twice has large as those distributed at year $Y+1$,
and have devalued by only 2\% in the meantime.
Thus, the users who participated at $Y$,
and held onto most of their basic incomes,
see a near-doubling of their account balances
as measured against the basic incomes distributed at $Y+1$.

Whether this effect is good or bad depends on goals and tradeoffs.
On this upside,
this effect creates a potentially strong early adopters' reward,
and incentivizes those early adopters to help sign up more users,
since the reward materializes only if and to the extent user population grows.
Since it seems both unlikely and risky in practice
to try to introduce a currency as unconventional as \coin
all at once in any large existing society,
its success will ultimately depend on growth based on voluntary adoption,
and appropriate early-adoption incentives may be essential
to the currency achieving critical mass.

On the other hand,
one clear potential risk is that of \coin appearing like a Ponzi scheme.
\xxx{citation?}
It is not, because the early-adoption rewards are both fully transparent
and gracefully self-limiting.
Anyone can trivially calculate the remaining potential early-adoption reward --
the factor their first basic income distribution could be multiplied by
if they join now --
by simply dividing the total population of the world
(or of those countries that might plausibly allow \coin to be widely adopted)
by the current user population.
If and when \coin is successful and nears saturation
among some relevant population of potential users,
the early-adoption reward factor gradually and smoothly
drops to one, \ie, no reward.
Conveniently, that is also precisely the point at which adoption incentives
are no longer needed.

A secondary risk is that his early-adoption reward
is likely to encourage hoarding and speculative investment in the short term,
in the same way we have observed in other cryptocurrencies such as Bitcoin.
This short-term effect is thus precisely the opposite of what
Gesell, Keynes, or we, would probably want in a stable ``working'' currency.
However, in contrast with classical macroeconomic economics
in which stability is {\em the} primary goal,
for us stability is only {\em one} of our goals
and not necessarily the most important in the short term,
as discussed earlier in Section~\ref{sec:money-principles}.
We are willing to sacrifice some short-term stability
in the interest of enabling adoption of a monetary system
that may promise to be much more fundamentally sound, sustainable,
and ultimately {\em stable},
in the long term.
Like a pilot caught in the eye of a growing hurricane,
we find it far preferable to tolerate
the bumpy and difficult ride that might be necessary to get out,
than to circle in place in pursuit of short-term stability alone
until we run out of fuel and crash.


\section{Property and Ownership in a Democratic Economy}
\label{sec:property}

{\em In preparation.}


\section{Securely Democratizing Information}

{\em In preparation.}

\xxx{

No democracy can be secure
unless its citizens have secure sources of accurate, unbiased information.
If an external attacker such as a resourceful foreign power
can mislead, distort, or even merely cast doubt on information sources,
or if an internal special-interest or ideological tribe can do the same,
then the democracy is vulnerable to hijacking and being led in directions
that represent neither nor the common good.
This section, therefore,
focuses on defining principles and potential technologies
to secure a democratic society's information sources in the digital age.

Principles:
\begin{itemize}
\item	Freedom of expression, anonymity
\item	Accountability, trolling-resistance, blockability
\item	Time and attention scarcity
\item	Fairness, resistance to jamming/flooding/trolling/loud voices.
\item	Personalisation
\item	Perspective breadth.
	Avoid echo chambers, polarization.
	Balance interest/proximity focus vs broad focus.
\item	Timescales
\item	Reputation, attention economy
\end{itemize}

attention economy, news

\subsection{Freedom of Expression}

Anonymity with accountability

\subsection{Personalization and Breadth}

Locality metrics: social network locality,
topic locality (\eg, hashtag), geographic locality

\subsection{Information Timescales}

\subsection{Discussion Localization}

Interest-based and representative (sample-based, \eg, deliberative polls)

\subsection{Attention Economics}

}

\section{Decision-Making}
\label{sec:decis}

{\em In preparation.}

\xxx{
Principles:
\begin{itemize}
\item	Scalable participation
\item	Freedom of type and level of participation
\item	Encourage consensus and cooperation, avoid polarization
\item	Ensure minority positions neither ignored nor overly empowered
\end{itemize}
}

\xxx{ Experimental measures of success:
refer to Steenbergen Discourse Quality Index

read: Habermas discourse ethics
(1981, 1990, 1991, 1992, 1995, 1996)

rules and procedures of the discourse should be
open for discussion
(Cohen, 1989; Habermas, 1992, 370–372; Chambers, 1995; Benhabib, 1996)

one could also create factor scores to form the scale (Gorsuch, 1983)

Can we create a deliberative forum with peer-review-based DQI coding?
}

\section{Innovation and Self-Evolution}
\label{sec:evo}

{\em In preparation.}

\section{Related Work}

{\em In preparation.}

\xxx{



}

\xxx{
highly relevant:
\href{https://arxiv.org/pdf/0705.4110.pdf}{Optimizing Scrip Systems: Efficiency, Crashes, Hoarders, and Altruists}
}

\section{Conclusion}

{\em In preparation.}

\com{

\subsection{Acknowledgments}

Landemore, Alex Lipton, Gun Sirer, ...
Misiek, Hanna?











}

\bibliography{soc,sec}
\bibliographystyle{plain}

\end{document}